\documentclass[twocolumn,showpacs,preprintnumbers,amsmath,amssymb,superscriptaddress]{revtex4-1}
\usepackage{graphicx}

\newcommand{\gsim}{\;\rlap{\lower 3.5 pt \hbox{$\mathchar \sim$}} \raise 1pt
 \hbox {$>$}\;}
\newcommand{\lsim}{\;\rlap{\lower 3.5 pt \hbox{$\mathchar \sim$}} \raise 1pt
 \hbox {$<$}\;}

\usepackage{color}

\allowdisplaybreaks

\begin{document}

\title{Leptonic decay of the Upsilon(1S) meson at third order in QCD}

\author{Martin Beneke}
\affiliation{Physik Department T31, James-Franck-Stra\ss{}e~1,
    Technische Universit\"at M\"unchen, 85748 Garching, Germany}
\affiliation{Institut f\"ur Theoretische Teilchenphysik und Kosmologie, RWTH
  Aachen, 52056 Aachen, Germany}

\author{Yuichiro Kiyo}
\affiliation{Department of Physics, Juntendo University, Inzai, Chiba, Japan}

\author{Peter Marquard}
\affiliation{Deutsches Elektronen Synchrotron DESY,
      Platanenallee 6, 15738 Zeuthen, Germany}

\author{\\Alexander Penin}
\affiliation{Department of Physics, 
 University of Alberta,
  Edmonton AB T6G 2J1, Canada}

\author{Jan Piclum}
\affiliation{Physik Department T31, James-Franck-Stra\ss{}e~1,
    Technische Universit\"at M\"unchen, 85748 Garching, Germany}
\affiliation{Institut f\"ur Theoretische Teilchenphysik und Kosmologie, RWTH
  Aachen, 52056 Aachen, Germany}

\author{Dirk Seidel}
\affiliation{Theoretische Physik 1, Universit\"at Siegen, 57068 Siegen, Germany}

\author{Matthias Steinhauser}
\affiliation{Institut f{\"u}r Theoretische Teilchenphysik, Karlsruhe
  Institute of Technology (KIT), 76128 Karlsruhe, Germany}

\date{January 13, 2014}

\begin{abstract}
\noindent 
  We present the complete next-to-next-to-next-to-leading order 
  short-distance and bound-state QCD correction 
  to  the leptonic decay rate $\Gamma(\Upsilon(1S)\to \ell^+\ell^-)$ of the 
  lowest-lying spin-1 bottomonium state. The perturbative QCD prediction is 
  compared to the measurement  $\Gamma(\Upsilon(1S)\to e^+e^-) = 
  1.340(18)$~keV.
\end{abstract}

\preprint{ALBERTA-THY-01-14,
  DESY-13-254,
  LPN14-004,  SFB/CPP-14-04,
  SI-HEP-2014-02,
  TTK-14-02,
  TTP14-003}
\preprint{TUM-HEP-923/14}
 
\maketitle

Bound states of a heavy quark and antiquark provide an ideal
laboratory to study non-relativistic quantum chromodynamics
(NRQCD). The bound-state dynamics is characterized by three scales,
the mass of the heavy quark (hard scale), $m$, its typical momentum
(soft scale), $mv$, and energy (ultrasoft scale),
$mv^2$. Here $v\sim\alpha_s(mv)$ is the velocity of the quark in the
bound state and $\alpha_s$ the strong coupling.
The theoretical description of heavy-quark bound states
uses the fact that the different scales are well-separated since the
velocity is small. This allows to construct a series of effective
theories by integrating out the larger scales. Starting from QCD, the
first step is to integrate out the hard modes to obtain
NRQCD~\cite{Thacker:1990bm,Lepage:1992tx,Bodwin:1994jh}. The second step is to integrate out potential and soft gluons and soft light quarks, 
leading to potential NRQCD
(PNRQCD)~\cite{Pineda:1997bj}. PNRQCD
contains only potential heavy quarks, whose energy and momentum are of
order $mv^2$ and $mv$, respectively, and ultrasoft gluons and light quarks.

A ``classical'' application of NRQCD is the prediction of the decay rate
of heavy-quark bound states into leptons. The simplest such system is the
$\Upsilon(1S)$ meson, the lowest-lying spin-triplet bound state of a bottom
quark and antiquark.  To next-to-next-to-next-to-leading order accuracy (N$^3$LO) 
the decay rate can be computed with the help of the formula~\cite{Beneke:2007gj}
\begin{eqnarray}
  \lefteqn{\Gamma(\Upsilon(1S)\to \ell^+\ell^-)}&& \nonumber \\
  &=& \frac{4\pi\alpha^2}{9m_b^2}
  \left|\psi_1(0)\right|^2
  c_v \left[c_v - \frac{E_1}{m_b} \left(c_v+\frac{d_v}{3}\right) + 
  \ldots\right] \,,\quad
\label{eq:master}
\end{eqnarray}
with $\alpha$ being the fine structure constant
and $m_b$ the bottom-quark pole mass.
$c_v$ and $d_v$ are matching constants of leading and sub-leading 
$b\bar b$ currents in NRQCD, and $\psi_1(0)$ is the 
wave function of the $(b\bar{b})$ system at the origin,
which at leading order is given by
\begin{eqnarray}
  \left|\psi_1^{\rm LO}(0)\right|^2 &=& \frac{8  m_b^3\alpha_s^3}{27 \pi}
  \,.
\end{eqnarray}
The mass of the $\Upsilon(1S)$ is $M_{\Upsilon(1S)} = 2 m_b+E_1$, and the 
perturbative part of the binding energy $E_1$ is given  at leading order 
by $E_1^{p,\rm LO}=-(4m_b\alpha_s^2)/9$.

In the following we assume that the bound-state dynamics of the 
$\Upsilon(1S)$ state is governed by weak coupling, which formally 
requires that the ultrasoft scale $m_b v^2$ is large compared to 
the strong interaction scale $\Lambda$. It is generally believed that 
this is a reasonable assumption for the lowest-lying $1S$ state, 
but not for the higher states, which, though more non-relativistic, 
are too large to be considered as bound states dominated by the 
colour-Coulomb interaction. Even for the $1S$ state the assumption 
$m_b v^2 \gg \Lambda$ is questionable. In fact, the leptonic decay 
that we consider in this Letter should be considered as one of the 
crucial tests of perturbative QCD bound-state dynamics, when all three 
scales (hard, soft, ultrasoft) are relevant to the problem. The more 
recent analyses of the leptonic $\Upsilon(1S)$ decay are based on 
next-to-leading order QCD together with non-perturbative condensate 
corrections~\cite{Pineda:1996uk}, or second-order QCD without 
non-perturbative corrections~\cite{Beneke:1999fe}, and both fail to 
describe the measured decay width accurately. The problem arises 
from large uncertainties in the perturbative and non-perturbative 
corrections. We address both issues in this Letter.

Recently the last missing ingredients for a complete N$^3$LO evaluation of
$\Gamma(\Upsilon(1S)\to \ell^+\ell^-)$ have been computed, which allow 
us to reconsider the problem with unprecedented accuracy:
The gluonic three-loop contributions to $c_v$ have been evaluated in
Ref.~\cite{MPSS}, and third-order corrections to the wave function at the 
origin induced by single- and double-potential insertions and ultrasoft gluon 
exchange are computed in
Ref.~\cite{Beneke:2007gj,Beneke:2007pj,BenKiy_I,BenKiy_II}. 
Furthermore, the still missing two-loop ${\cal O}(\epsilon)$
term of the $d=4-2\epsilon$ dimensional matching 
coefficient of the $1/(m_b r^2)$ PNRQCD potential is given in the Appendix.

Thus, we are now in the position to compute the decay rate of the
$\Upsilon(1S)$ meson into a lepton pair to third order in perturbation
theory. The following results apply to the cases $\ell=e,\mu$, where 
the lepton mass can be neglected. 
Expanding out all factors of Eq.~(\ref{eq:master}) in $\alpha_s\equiv 
\alpha_s(\mu)$, where $\mu$ denotes the renormalization scale in the 
$\overline{\rm MS}$ scheme, we obtain 
\begin{eqnarray}
  \lefteqn{\Gamma(\Upsilon(1S)\to \ell^+\ell^-)|_{\mathrm{pole}}} && 
\nonumber \\
  &=& 
 \frac{2^5\alpha^2\alpha_s^3m_b}{3^5}\,\Big[ 1+
\alpha_s \left(-2.003+3.979\,L\right) 
\nonumber \\[0.1cm]
&& +\, 
\alpha_s^2 \left(9.05-7.44 \,\ln\alpha_s -13.95 \, L 
+ 10.55 \,L^2\right)
\nonumber\\[0.2cm]
&& +\, \alpha_s^3\left(-0.91
+ 4.78_{a_3} 
+ 22.07_{b_2\epsilon} 
+ 30.22_{c_f} 
\right.
\nonumber\\
&& \hspace*{0.7cm} - \,134.8(1)_{c_g}
- 14.33 \,\ln\alpha_s 
- 17.36 \,\ln^2\alpha_s 
\nonumber\\
&& \hspace*{0.7cm} 
+\,(62.08- 49.32 \,\ln\alpha_s) \,L 
- 55.08 \,L^2
\nonumber\\
&& \hspace*{0.7cm} 
\left.+ \,23.33 \,L^3\right)+ {\cal O}(\alpha_s^4)\,
\Big]
\label{eq:gampole1}\\
&=& \frac{2^5\alpha^2\alpha_s^3m_b}{3^5}
  \left[1 + 1.166\alpha_s + 15.2\alpha_s^2
    +\left(66.5
    + 4.8_{a_3} 
    \right.\right.\nonumber\\&&\left.\left.\mbox{}
      + 22.1_{b_2\epsilon} 
      + 30.2_{c_f}
      - 134.8(1)_{c_g}\right)
    \alpha_s^3
  + {\cal O}(\alpha_s^4)\right]  \nonumber \\[0.2cm]
  &=& \frac{2^5\alpha^2\alpha_s^3m_b}{3^5}
  \left[1+ 0.28 + 0.88 - 0.16 \right] \nonumber \\[0.2cm]
  &=& [1.04\pm 0.04(\alpha_s){}^{+0.02}_{-0.15}(\mu)]~\mbox{keV} \,,
\label{eq:gampole3} 
\end{eqnarray}
where $L = \ln\,(\mu/(m_bC_F\alpha_s))$ with $C_F=4/3$.
The subscripts indicate the contribution from the (scale independent)
coefficients of the three-loop static potential ($a_3$), the ${\cal
  O}(\epsilon)$ term of the $1/(m_b r^2)$ potential ($b_2\epsilon$) and the
fermionic and bosonic contribution of the three-loop matching coefficient
($c_f$ and $c_g$).
The uncertainty due to the limited precision of the latter is given in
parentheses.
The contribution from the  ${\cal O}(\epsilon)$ terms of the $1/(m_b^2
r^3)$ potentials is not made explicit.

For the numerical evaluation after Eq.~(\ref{eq:gampole1}) we use 
$\alpha(2m_b)=1/132.3$~\cite{Jegerlehner:2011mw}, $\alpha_s(M_Z)=0.1184(10)$
and the renormalization scale $\mu=3.5$~GeV. We use the program {\tt
  RunDec}~\cite{Chetyrkin:2000yt} to evolve the 
coupling in the four-loop approximation such that 
$\alpha_s(3.5\,\mbox{GeV})=0.2411$ and 
to compute the pole mass $m_b=4.911$~GeV in the 3-loop 
approximation from the $\overline{\mathrm{MS}}$ 
value $\bar{m}_b(\bar{m}_b)=4.163(16)$~GeV given in
Ref.~\cite{Chetyrkin:2009fv}. The scale uncertainty in Eq.~(\ref{eq:gampole3}) 
is computed from the maximum and minimum value of the width within 
the range $\mu\in [3,10]\,\mbox{GeV}$ (see discussion below).
Note that the uncertainty induced by the bottom-quark mass is below 1 per
mille and can thus be neglected.
However, this does not take into account the uncertainty due to the
perturbative instability of the pole mass.

We can avoid the computation of the pole mass
by going to the potential-subtracted (PS) 
mass scheme~\cite{Beneke:1998rk}. In computing the PS mass from 
the $\overline{\rm MS}$ mass $\bar{m}_b(\bar{m}_b)=4.163$~GeV, 
we combine (for $n=1,2,3$) the $n$-loop correction to the 
$\overline{\rm MS}$-pole-mass
relation with the $(n-1)$-loop correction to the Coulomb potential 
in the pole-PS-mass relation and find $m_b^{\mathrm{PS}}\equiv 
m_b^{\mathrm{PS}}(\mu_f=2\,\mbox{GeV})=4.484$~GeV. We then eliminate 
$m_b$ in Eq.~(\ref{eq:gampole1}) by replacing $m_b = m_b^{\mathrm{PS}} 
+ \delta m$ and expand systematically in $\alpha_s$ to obtain
\begin{eqnarray}
  \lefteqn{\Gamma(\Upsilon(1S)\to \ell^+\ell^-)|_{\mathrm{PS}}} && 
\nonumber \\[0.1cm]
  &=& \Gamma(\Upsilon(1S)\to \ell^+\ell^-)|_{\mathrm{pole}, 
m_b\to m_b^{\mathrm{PS}}}
\nonumber \\
  && +\, 
 \frac{2^5\alpha^2\alpha_s^3m_b^{\mathrm{PS}}}{3^5}\,x_f\,\Big[
0.42 \alpha_s^2 
+ \alpha_s^3\,\big(-1.78 + 0.28\, L_f
\nonumber\\
&& \hspace*{0.5cm} 
+ 1.69\,L \big)+ {\cal O}(\alpha_s^4)\,
\Big]\\
  &=& \frac{2^5\alpha^2\alpha_s^3m_b^{\mathrm{PS}}}{3^5}
  \left[1 + 1.528\alpha_s + 16.3\alpha_s^2
    +\left(74.7
    + 4.8_{a_3} 
    \right.\right.\nonumber\\&&\left.\left.\mbox{}
      + 22.1_{b_2\epsilon} 
      + 30.2_{c_f}
      - 134.8(1)_{c_g}\right)
    \alpha_s^3
  + {\cal O}(\alpha_s^4) \right] 
\nonumber\\
  &=& \frac{2^5\alpha^2\alpha_s^3m_b^{\mathrm{PS}}}{3^5}
  \left[1+ 0.37 + 0.95 - 0.04 \right] 
\nonumber\\
  &=& [1.08\pm 0.05(\alpha_s){}^{+0.01}_{-0.20}(\mu)]~\mbox{keV}
  \,,
\end{eqnarray}
with $x_f = \mu_f/(m_b^{\mathrm{PS}}\,\alpha_s)$ and $L_f=\ln(\mu^2/\mu_f^2)$.
The pattern of the series is essentially the same in both schemes.
The NNLO corrections are very large~\cite{Beneke:1999fe}, but we find
only moderate
corrections at N$^3$LO. Together with the improved scale dependence at
third order discussed below, this may be an indication that perturbative 
corrections beyond the third order are small.

\begin{figure}[t]
  \includegraphics[width=\columnwidth]{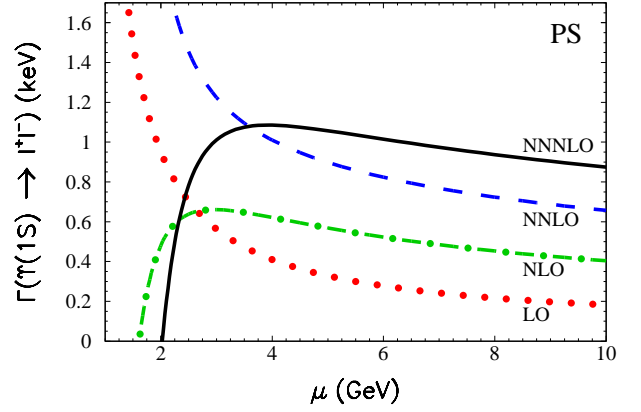}
  \caption{\label{fig::Gamma} The decay rate in the PS scheme as a
    function of the renormalization scale $\mu$. Dotted (red),
    dash-dotted (green), short-dashed (blue) and solid (black) lines
    correspond to LO, NLO, NNLO and N$^3$LO prediction.} 
\end{figure}

In Fig.~\ref{fig::Gamma} we show the decay rate
$\Gamma(\Upsilon(1S)\to \ell^+\ell^-)$ in the PS
scheme
as a function of the renormalization scale including successively
higher orders. Very similar results are obtained for the pole scheme.
For small scales no convergence is observed and for
values close to the soft scale $\mu_s = m_b \alpha_s(\mu_s) C_F
\approx 2.0$~GeV  there are big differences between subsequent
perturbative orders. It is interesting to note that  for $\mu\gsim
3$~GeV the N$^3$LO prediction becomes quite flat and furthermore only
shows a small deviation from the NNLO curve. We take this as evidence 
that perturbative computations of Coulomb bound states in 
QCD are better behaved when the scale is taken somewhat larger 
than the naive estimate of the soft scale, as already observed 
in Ref.~\cite{Beneke:2005hg}, and vary $\mu$ between $3\,$GeV and 
$10\,$GeV to compute the scale uncertainty.

\begin{figure}[t]
  \includegraphics[width=\columnwidth]{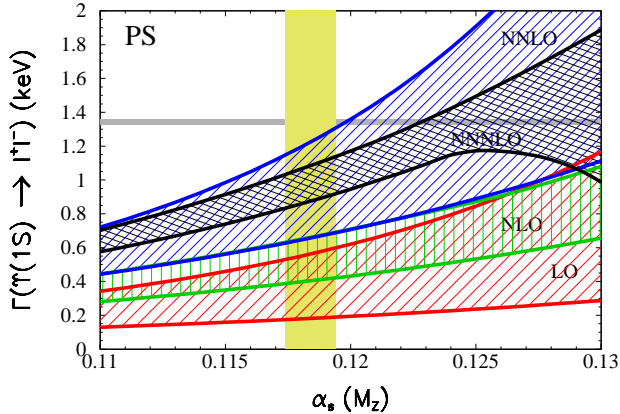}
  \caption{\label{fig::GamAsPS}The decay rate as a function of $\alpha_s(M_Z)$
    at LO (red, bottom), 
    NLO (green, middle), NNLO (blue, top), 
    and N$^3$LO (black, inner top band).  The
    bands denote the variation of $\mu$ between 3~GeV and 10~GeV. The
    horizontal bar denotes the experimental value, while the vertical bar
    denotes the world average of the strong coupling constant,
    $\alpha_s(M_Z)=0.1184(10)$.}
\end{figure}

Compared to the experimental value $\Gamma(\Upsilon(1S)\to
e^+e^-)|_{\mathrm{exp}} = 1.340(18)$~keV~\cite{Beringer:1900zz}, the 
third-order perturbative result is about $30\%$ too low. The 
discrepancy remains substantial even when including the theoretical 
uncertainty. Note, however, 
that the decay rate depends on the value of $\alpha_s$ 
to a high power. In Fig.~\ref{fig::GamAsPS} we therefore show
the decay rate and its scale dependence as a function of
$\alpha_s(M_Z)$ at LO, NLO, NNLO, and N$^3$LO in the PS scheme. The plot
shows good convergence of the perturbative series up to
$\alpha_s(M_Z)\approx 0.122$, with the N$^3$LO band completely inside
the NNLO one. However, the third-order result is always below the
experimental value up to this point. 

Since the perturbative contributions seems to be well under control at 
third order, a possible explanation for the difference between the 
experimental and the perturbative value is a sizable non-perturbative 
contribution. This is not implausible, since the scale of ultrasoft 
gluons is close to the strong-interaction scale for the $\Upsilon(1S)$ 
meson. The contribution
to the wave function at the origin due to the gluon condensate has been
evaluated in Refs.~\cite{Pineda:1996uk,Voloshin:1979uv}. It takes the 
form
\begin{equation}
\delta_{\mathrm{np}}|\psi_1(0)|^2 = |\psi_1^{\rm LO}(0)|^2\, 
\times 17.54 \pi^2 K\,,
\label{eq:gamnonpert}
\end{equation}
where
\begin{equation}
K = \frac{\langle \frac{\alpha_s}{\pi} G^2 \rangle}{m_b^4(\alpha_s
  C_F)^6}
\label{eq::def_K}
\end{equation}
is the dimensionless number that controls the relative size of the 
gluon condensate contribution.
Using $\langle \frac{\alpha_s}{\pi}G^2\rangle =
0.012$~GeV$^4$~\cite{Shifman:1978by}  and 
$\alpha_s(3.5\,\mbox{GeV})$, its contribution to
the decay rate evaluates to 
$\delta_{\mathrm{np}}\Gamma_{\ell\ell}(\Upsilon(1S)) = 1.67$~keV 
in the pole mass scheme and 2.20~keV in the PS mass
scheme, far in excess of the missing $0.26\,$keV.
There is a large uncertainty in these estimates, since the value of 
the gluon condensate is very uncertain and the scale of $\alpha_s$ in the 
denominator is undetermined. For example, if we adopt 
the strategy of Ref.~\cite{Pineda:1996uk} and replace $\alpha_s$ 
in the denominator of Eq.~(\ref{eq::def_K}) by a coupling 
$\tilde{\alpha}_s$ related to the coefficient of the Coulomb potential 
at the scale $\mu = 1\,$GeV, the above numbers change to 
0.06~keV (pole scheme) and 0.08~keV (PS scheme), respectively. 
Moreover, depending on the choice for the strong coupling 
in Eq.~(\ref{eq::def_K}), one either concludes from the 
size of the dimension-6 condensate contribution, also computed in 
Ref.~\cite{Pineda:1996uk}, that 
the condensate expansion is not convergent, or, to the contrary, well behaved.
Hence, no reliable estimate of the 
leptonic decay width can be obtained by this procedure. 

Additional insight on $\delta_{\mathrm{np}}|\psi_1(0)|^2$ can be 
obtained from the mass of the 
$\Upsilon(1S)$ state, which we can write as 
\begin{equation}
M_{\Upsilon(1S)} = 2 m_b + E_1^{\rm p} + \frac{624\pi^2}{425} 
\,m_b(\alpha_s C_F)^2\,K\,, 
\label{eq:M1Snonpert}
\end{equation}
where $E_1^{\rm p}$ is the perturbative contribution to the bound-state 
energy, which is also known to the third order in 
QCD~\cite{Kniehl:2002br,Penin:2002zv,Beneke:2005hg}, and $K$ is the gluon
condensate  
correction from Refs.~\cite{Pineda:1996uk,Voloshin:1979uv,Leutwyler:1980tn}.
For the following analysis, it is mandatory to work with the PS 
scheme to achieve a reliable perturbative expansion of 
$E_1^p$ (cf.~Ref.~\cite{Beneke:2005hg}, Eq.~(38)). A direct determination 
of $m_b^{\rm PS}$ from the  $\Upsilon(1S)$ mass at third order, but 
excluding the non-perturbative contribution, 
gives  $m_b^{\rm PS} = 4.57\,\mbox{GeV}$~\cite{Beneke:2005hg} 
(the central scale $\mu\approx 2\,$GeV is used in this 
reference), which is larger 
than the value $4.48\,\mbox{GeV}$ obtained above from the most accurate 
determinations of the $\overline{\rm MS}$ bottom-quark mass. 
This suggests that there is a non-negligible 
non-perturbative contribution $\delta M_{\Upsilon(1S)}^{\rm np}$ 
to the $\Upsilon(1S)$ mass. Repeating 
the analysis of Ref.~\cite{Beneke:2005hg} with our parameters, we
find
\begin{eqnarray}
\delta M_{\Upsilon(1S)}^{\rm np} &\equiv& M_{\Upsilon(1S)} - 
(2 m_b^{\rm PS} + E_1^{\rm p, PS}) \nonumber\\
 &\approx& [125 \pm 16(\alpha_s) \pm 34 (m_b)
 {}^{+10}_{-25}(\mu)]\,\mbox{MeV}\,, \qquad
\label{dMnpEstimate}
\end{eqnarray} 
where $E_1^{\rm p, PS} = 2 m_b -2 m_b^{\rm PS} + E_1^p$. 
This estimate is considerably larger than the value 
$\delta M_{\Upsilon(1S)}^{\rm np} \approx 15\,$MeV given in 
Ref.~\cite{Pineda:1996uk} based on the condensate expansion, 
and relies only on the accurate input value for the 
bottom $\overline{\rm MS}$ mass and the convergence of 
the perturbative expansion of the binding energy in the 
PS scheme~\cite{Beneke:2005hg}. 

Eq.~(\ref{dMnpEstimate}) neglects the mass of the charm quark. 
The effect of a finite mass $m_c=1.4\,$GeV is easily 
computed at ${\cal O}(\alpha_s^2)$ and reduces 
$\delta M_{\Upsilon(1S)}^{\rm np}$ by $12\,$MeV for given 
$\overline{\rm MS}$ bottom-quark mass. Including an estimate 
of the next order from Ref.~\cite{Hoang:2000fm}, we therefore 
subtract $(20\pm 10)\,$MeV from Eq.~(\ref{dMnpEstimate}). 
Comparing Eq.~(\ref{eq:M1Snonpert}) 
to Eq.~(\ref{eq:gamnonpert}), we find the relation 
\begin{eqnarray}
&& \delta_{\mathrm{np}}\Gamma_{\ell\ell}(\Upsilon(1S)) = 
    \frac{4\alpha^2\alpha_s}{9}\,\frac{17.54\times 425}{3744}\,
    \delta M_{\Upsilon(1S)}^{\rm np}
\label{eq:relMGam}\\[0.1cm]
&& \approx \,
[1.28 {}^{+0.17}_{-0.18}(\alpha_s) \pm 0.42 (m_b) {}^{+0.20}_{-0.57}
(\mu) \pm 0.12 (m_c)]\,\mbox{keV}\,.
\nonumber
\end{eqnarray}
The numerical result is closer to the larger values obtained 
in our previous estimates. It must, however, be taken with a grain 
of salt, since for such large values the condensate expansion 
is not convergent. The different sub-leading dimension-6 corrections 
to $\delta_{\mathrm{np}}\Gamma_{\ell\ell}(\Upsilon(1S))$ and 
$\delta M_{\Upsilon(1S)}^{\rm np}$ then invalidate the simple 
relation (\ref{eq:relMGam}) and once again preclude a 
reliable estimate of the non-perturbative part of the 
leptonic decay width. We should emphasize that this conclusion 
depends strongly on the state-of-the-art value 
$\bar{m}_b(\bar{m}_b)=4.163(16)$~GeV of the $\overline{\mathrm{MS}}$ 
mass~\cite{Chetyrkin:2009fv}. If the mass were only $40\,$MeV larger, we 
would find 
$\delta_{\mathrm{np}}\Gamma_{\ell\ell}(\Upsilon(1S)) \approx 0.3\,$keV from 
Eq.~(\ref{eq:relMGam}) and simultaneously conclude that the 
condensate expansion is well behaved.

In summary, we have computed the third-order
correction to the decay rate $\Gamma(\Upsilon(1S) \to l^+l^-)$. 
This is the first third-order QCD bound-state calculation, 
where both short- and long-distance effects are important. 
Both in the pole and potential subtracted scheme
the N$^3$LO corrections are negative and amount to about
$-16\%$ and $-4\%$, respectively. The perturbative uncertainty 
that constituted the main limitation of previous analyses is 
thus mostly removed. We find that the leptonic decay width is mostly 
perturbative; the perturbative contribution 
amounts to roughly 70\% of the measured value. 
The new third-order contribution is crucial to ascertain this 
conclusion.
We further considered several estimates of non-perturbative effects 
based on the condensate expansion, including a relation 
to the mass of the $\Upsilon(1S)$ state. Unfortunately, 
the situation is ambiguous and no clear conclusion on the size 
of non-perturbative effects could be drawn. Whether a full 
quantitative, theoretical understanding of the leptonic decay width 
can be achieved therefore remains an open question. We note, however, 
that this conclusion relies on the precise value of 
the bottom   $\overline{\rm MS}$ quark mass.

\section*{Acknowledgments}

\noindent 
This work has
been supported by the DFG SFB/TR~9
``Com\-pu\-ter\-gest\"utzte Theoretische Teil\-chen\-physik,'' the Gottfried
Wilhelm Leibniz programme of the Deutsche Forschungsgemeinschaft (DFG),
the DFG cluster of excellence ``Origin and Structure of the
Universe,'' and the EU Network {\sf LHCPHENOnet} PITN-GA-2010-264564.
The work of A.P. is supported in part by NSERC and the Alberta Ingenuity
Foundation.

\section*{\boldmath Appendix: Two-loop ${\cal O}(\epsilon)$ term of $1/(m_b r^2)$
  potential} 

\noindent 
In this Appendix we present the result for the two-loop ${\cal O}(\epsilon)$
term of the matching coefficient of the $1/(m_b r^2)$ potential. This is most
easily achieved by replacing the quantity $b_2$ in Eq.~(6) of
Ref.~\cite{Kniehl:2001ju} by $b_2 + \epsilon b_2^\epsilon$. Then
$b_2^\epsilon$ reads
\begin{eqnarray}
  b_2^\epsilon &=&
  C_FC_A\left(-\frac{631}{108} - \frac{15\pi^2}{16} + \frac{65\ln2}{9}
    - \frac{8\ln^2 2}{3}\right)
  \nonumber\\&&\mbox{}
  + C_A^2 \left( -\frac{1451}{216} - \frac{161\pi^2}{72} 
    - \frac{101\ln2}{18} - \frac{4\ln^2 2}{3}\right)
  \nonumber\\&&\mbox{}
  + C_A T n_l \left(\frac{115}{54} + \frac{5\pi^2}{18} + \frac{49\ln 2}{18}\right)
  \nonumber\\&&\mbox{}
  + C_F T n_l \left(\frac{17}{27} - \frac{11\pi^2}{36} - \frac{4\ln 2}{9}\right)
  \,.
\end{eqnarray}

\end{document}